\documentclass[aps,prd,twocolumn,superscriptaddress,showpacs]{revtex4}
\usepackage{graphicx}
\usepackage{dcolumn}
\usepackage{bm}
\usepackage{natbib}

\newcommand{\be}{\begin{equation}}
\newcommand{\ee}{\end{equation}}
\newcommand{\bea}{\begin{eqnarray}}
\newcommand{\eea}{\end{eqnarray}}

\newcommand{\WMAP}{{\slshape WMAP~}}
\newcommand{\LCDM}{$ \Lambda $CDM~}

\newcommand{\HEALPIX}{{\slshape HEALPIX~}}
\newcommand{\CMBFAST}{\texttt {cmbfast~}}

\begin{document}
\title{A~high~redshift~detection~of~the~integrated~Sachs--Wolfe~effect}

\author {Tommaso~Giannantonio}

\affiliation {Institute of Cosmology and Gravitation, University
of Portsmouth, Mercantile House, Hampshire Terrace, Portsmouth PO1 2EG, UK}

\author {Robert~G.~Crittenden}

\affiliation {Institute of Cosmology and Gravitation, University
of Portsmouth, Mercantile House, Hampshire Terrace, Portsmouth PO1 2EG, UK}

\author {Robert~C.~Nichol}

\affiliation {Institute of Cosmology and Gravitation, University
of Portsmouth, Mercantile House, Hampshire Terrace, Portsmouth PO1 2EG, UK}

\author {Ryan~Scranton} 
\affiliation {Department of Physics and Astronomy, University of Pittsburgh, 3941 O'Hara St., Pittsburgh, PA 15260, USA}

\author {Gordon~T.~Richards}
\affiliation {Department of Physics and Astronomy, The Johns Hopkins University, 3400 N Charles Street, Baltimore, MD 21218, USA}

\author {Adam~D.~Myers} 
\affiliation {Department of Astronomy, University of Illinois at Urbana--Champaign, Urbana, IL 61820, USA}

\author {Robert~J.~Brunner}
\affiliation {Department of Astronomy, University of Illinois at Urbana--Champaign, Urbana, IL 61820, USA}

\author {Alexander~G.~Gray} 
\affiliation {Georgia Institute of Technology, Atlanta, GA 30332, USA}

\author {Andrew~J.~Connolly}
\affiliation {Department of Physics and Astronomy, University of Pittsburgh, 3941 O'Hara St., Pittsburgh, PA 15260, USA}

\author {Donald P. Schneider}
\affiliation {Department of Astronomy and Astrophysics, The Pennsylvania State University, University
Park, PA 16802, USA}

\begin {abstract}
We present evidence of a large angle correlation between the cosmic microwave background measured by \WMAP and a catalog of  photometrically detected quasars from the SDSS.  The observed cross correlation is $0.30 \pm 0.14 \mu K$ at zero lag, with a shape consistent with that expected for correlations arising from the integrated Sachs--Wolfe effect.  The photometric redshifts of the quasars are centered at $z\sim 1.5$, making this the deepest survey in which such a correlation has been observed.  Assuming this correlation is due to the ISW effect, this constitutes the earliest evidence yet for dark energy and it can be used to constrain exotic dark energy models.  
\end {abstract}


\maketitle

\section {Introduction} \label {sec:intro}

There is growing evidence that the expansion rate of the Universe is accelerating, which is believed to 
be the result of an unknown `dark energy' which has come to dominate the present energy density.  This is 
supported by many measurements, particularly recent observations of anisotropies in the cosmic microwave background (CMB) anisotropies, such as  by the Wlikinson Microwave Anisotropy Probe (\WMAP) \cite{Bennett:2003bz,Hinshaw:2006ia}, together with the present accelerated expansion inferred from the measurement of the Hubble diagram of the Type Ia supernovae \cite{Riess:2004nr, Astier:2005qq}.  Understanding the fundamental nature of this dark energy, what it might be and how it has evolved, is one of the biggest challenges facing cosmologists. 

One way of probing the dark energy and its evolution is through the integrated Sachs--Wolfe (ISW) effect \cite{Sachs:1967er}, which produces anisotropies in the CMB relatively recently.  While most of the CMB anisotropies were generated near the last scattering surface at the moment of recombination, when the universe was 400 ky old ($ z \simeq 1100 $), additional anisotropies can be created later by gravitational interactions.  If the gravitational potential $ \Phi $ varies on large scales at some stage of the universe's evolution, the CMB photons will undergo an energy shift
\be \label {eq:isw}
\Theta \equiv \frac {\Delta T} {T} = -2 \int \dot {\Phi} d\tau,
\ee
where $ c = 1 $, $ \tau $ is the conformal time, the dot represents a conformal time derivative and the integration is along the line of sight of the photon. The ISW effect reflects the fact that a photon falling into a potential well will climb out at the same energy only if the gravitational potential itself is constant in time, and will otherwise receive an energy shift depending on the evolution of the potential. 
During the matter--dominated era, the gravitational potential remains constant and so $ \dot{\Phi} = 0 $, leading to no ISW effect.  However, if the Universe becomes dominated by curvature or dark energy, then new CMB anisotropies can be created. 

The observed features of the CMB anisotropy spectrum indicate that the Universe is very close to flat, so a detection of the ISW can help constrain the dark energy.  Unfortunately, the ISW anisotropies will appear as a small addition to the anisotropies arising at higher redshift; they cannot be easily distinguished because their signal is largest on very large scales where cosmic variance is also large.   One way to extract this signal is to correlate the entire CMB anisotropy map with some tracer of the dark matter distribution \cite{Crittenden:1995ak,Afshordi:2004kz,Peiris:2000kb}: the primary CMB anisotropies will not be correlated in any way with the matter overdensities we observe now, because all these structures have formed much later.

The precise large scale CMB maps provided by the \WMAP satellite have made observing such weak correlations possible.
Many groups have detected the correlations between the \WMAP CMB data and different tracers of the large scale structure,  with results generally consistent with the predictions of the integrated Sachs--Wolfe effect for a dark energy model. 
The correlations have been detected with X-ray surveys \cite{Boughn:2003yz}, radio galaxy surveys \cite{Boughn:2003yz, Nolta:2003uy}, infrared observations \cite{Afshordi:2003xu} as well as optical surveys like the APM and the Sloan Digital Sky Survey (SDSS) \cite{Fosalba:2003iy, Scranton:2003in, Fosalba:2003ge, Padmanabhan:2004fy, Cabre:2006qm} using both ordinary and luminous red galaxy samples.  These surveys have also probed much different redshifts, allowing us to see the evolution of the correlation. 

In this paper, we use the NBC--KDE quasar catalog \cite{Richards:2004cz, Myers:2005jk} from the SDSS.  Objects in the NBC--KDE catalog are at $\bar{z}\sim 1.5$, making this the highest redshift sample ever used to probe the ISW effect. We find a positive signal, suggesting that the dark energy behaves in a way compatible to the cosmological constant up to a redshift of 1.5. In the following we briefly describe the quasar catalog used (section \ref {sec:qso}), calculate its autocorrelation (section \ref{sec:acf}).  Next we perform the cross--correlation with the CMB map (section \ref {sec:corr}). Finally we describe the resulting constraints in section \ref {sec:resul}, followed by the conclusions.

\section {The Quasar catalog} \label {sec:qso}

\begin{figure}[htbp] 
\begin{center}
\includegraphics[angle=0,width=3.5in]{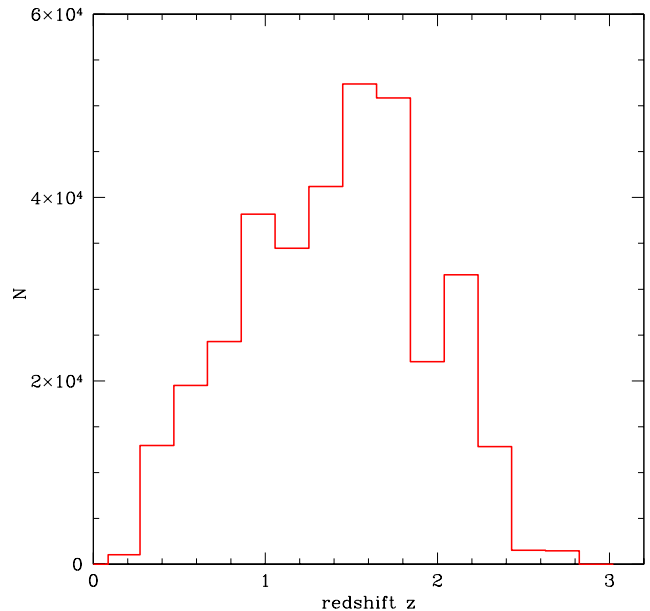}
\caption{Redshift distribution of the quasars.  A spline fit of this is used for the theoretical calculations.}
\label {fig:zeta}
\end{center}
\end{figure}

The quasar data was derived from SDSS DR4 \cite {Adelman-McCarthy:2005se, Fukugita:1996qt, Gunn:1998vh, Gunn:2006tw, Hogg:2001gc, Ivezic:2004bf, Lupton:1999pt, Pier:2002iq, Smith:2002pc, Stoughton:2002ae, York:2000gk}, using a nonparametric Bayes classifier method based on kernel density estimation (NBC--KDE) described in \cite {Richards:2004cz}. Briefly, this algorithm classifies quasars based on prior multi--color data on known quasars and stars, and is $ > 95 \% $ complete, with $ \sim 5 \% $ stellar contamination, to $ i = 21 $ \cite {Richards:2004cz, Myers:2005jk}.
The catalog contains $ N_q = 344,431 $ objects with photometric redshift between 0.1 and 2.7 (see Fig. \ref {fig:zeta}), covering two distinct regions of the northern hemisphere of the Galaxy plus three narrow stripes in the southern, covering a total area of 6,670 square degrees. 

The stellar contamination is a potentially important systematic.  Even if it does not contribute to the cross correlation 
(assuming the Galaxy has been cleaned from the CMB maps), it will still contribute to the quasar autocorrelation function.  The stellar spatial overdensities adds power on fairly large angular scales, which is difficult to explain with quasars alone.  Even a stellar contamination as small as $ 5 \% $ can produce significant angular overdensities in the $5-10^\circ$ range, where little contribution is expected from the quasars. For this reason, whenever modeling the expected behavior of our sample, we will assume that it is actually composed by a fraction $ k $ of stars and $ 1-k $ of quasars. In section III we will show the contamination is $ k = 0.05 \pm 0.01 $. 


\subsection*{Pixellation mask}

We are principally interested in the large scale correlations; to calculate these, we pixelize the quasar maps using the same {\slshape HEALPix} schema \cite {Gorski:2004by} used to pixelize the \WMAP maps.  We perform most correlations with a resolution parameter $ N_\mathrm{side} = 64 $, corresponding to $ N_\mathrm{pix} = $ 49,152 pixels of $ 0.92^{\circ}$ resolution. Due to the partial sky coverage of the survey, only $ 16 \% $ of these pixels actually contain sources.

Clearly the {\slshape HEALPix} pixelization will not exactly align with the SDSS regions where the quasars are observed.  Some edge pixels will thus be only partially filled and it is important to take these effects into account:  in the coarse pixelization described above, up to $ 20 \% $ of the pixels will be partially filled.  To account for such effects, we use a high--resolution ($ N_\mathrm{side}^{\mathrm{high}} = 512 $) pixelization to determine the mask of the actual sky coverage of DR4, which will determine the fraction of each edge pixel that is matching the observed area.  We base the mask on a random sample of galaxies in the DR4 database to ensure roughly uniform sampling in all directions.  (A much larger stellar sample would be needed because of the high concentration of sources close to the Galactic plane.)  By using a sufficiently large number of random galaxies ($ 5 \cdot 10^6 $) we can be sure to have good sampling when pixelized in the higher resolution. 
In the high--resolution map there are $ N_\mathrm{pix}^{\mathrm{high}} = 3 \cdot 10^6 $ total pixels, of which only $5 \cdot 10^5 $ cover the area of the survey, which means an average of 10 objects per pixel. 
In this way we estimate the coverage fraction of each low--resolution pixel, $ f_i $, as
\be
f_i = \frac {N_{\mathrm{mask}}^{\mathrm{high}}(i)} {64},
\ee
where $ {N_{\mathrm{mask}}^{\mathrm{high}}(i)} $ is the number of high--resolution pixels within the mask for each coarse pixel $ i $, and for these resolutions, there are $64$ high resolution pixels in each coarse pixel.  

We correct the maps by dividing the observed number of quasars in a coarse pixel by the fraction of the sky within the pixel that was observed, yielding $n_i/f_i$.  
For our correlation estimator, we down--weight such edge pixels by the fraction of sky they measure; this effectively accounts for the additional variance.  
A more conservative approach is to simply drop these edge pixels, ignoring all quasars in them (although at this resolution they contain roughly $ 20 \% $ of the catalog): we repeated our cross--correlation analysis using this schema and we found compatible results.

We use the higher resolution to calculate the average number of quasars per coarse pixel, $ \bar n $: this quantity is the total number of quasars divided by the total area of pixels covered by the survey in the higher resolution, rescaled to the pixel surface area in the lower resolution:
\be
\bar n = \frac{N_q}{N_\mathrm{mask}^\mathrm{high}} \times 64 ,
\ee
where $ N_\mathrm {mask}^{\mathrm {high}} = \sum_i N_\mathrm {mask}^{\mathrm {high}} (i)$ is the total number of higher resolution pixels within the mask.

\subsection* {Foregrounds} \label{ssec:sys}

There are a number of possible systematics in the catalog which could introduce errors resulting in a lack of completeness, bad redshift measurement or further stellar contamination; these could introduce artificial structures in the maps and contaminate the measurements.  We checked a number of these, including extinction by dust in our Galaxy, sky brightness, bright star obscuration and poor seeing in two different bands ($r$ and $g$).

The SDSS imaging data is obtained using drift--scanning, which produces long thin strips of data across the sky. Two adjacent strips are combined to make a stripe, which are then chopped into individual fields of dimension $ 10 \times 10 $ arcmins \cite{York:2000gk}. Clearly,  this observing strategy could introduce small correlations along a strip (or stripe), which could extend to very large angles (over 100 degrees) in the imaging data, e.g. systematic differences in the zero--point calibration of the photometry in each strip. Such photometric calibration uncertainties were recently explored by \cite{Padmanabhan:2006ku} and shown to be less than $ 2 \% $, consistent over the whole SDSS area. This is below other statistical (shot noise) and systematic (extinction, seeing) errors and therefore is not considered further here. We also note that the SDSS scanning strategy is not aligned with any cosmological or galactic signal and would therefore only introduce extra noise into our ISW detection rather than mimicing the signal.

While the extinction is a quantity measurable for each observed object, sky brightness, seeing and number of point sources are global quantities of each $ 10 \times 10 $ arcmins field of view. However, given an object we can find the foreground quantities associated to it through the ID number of its field of view and, because the fields are smaller ($ \sim 1 / 25 $) than our pixels, we can consider the distribution of all these quantities in each pixel in the same way.

\begin{figure}[htb] 
\begin{center}
\begin{tabular}{c c}
\begin{minipage}{0.5\linewidth}
\includegraphics[angle=0,width=1.0\linewidth]{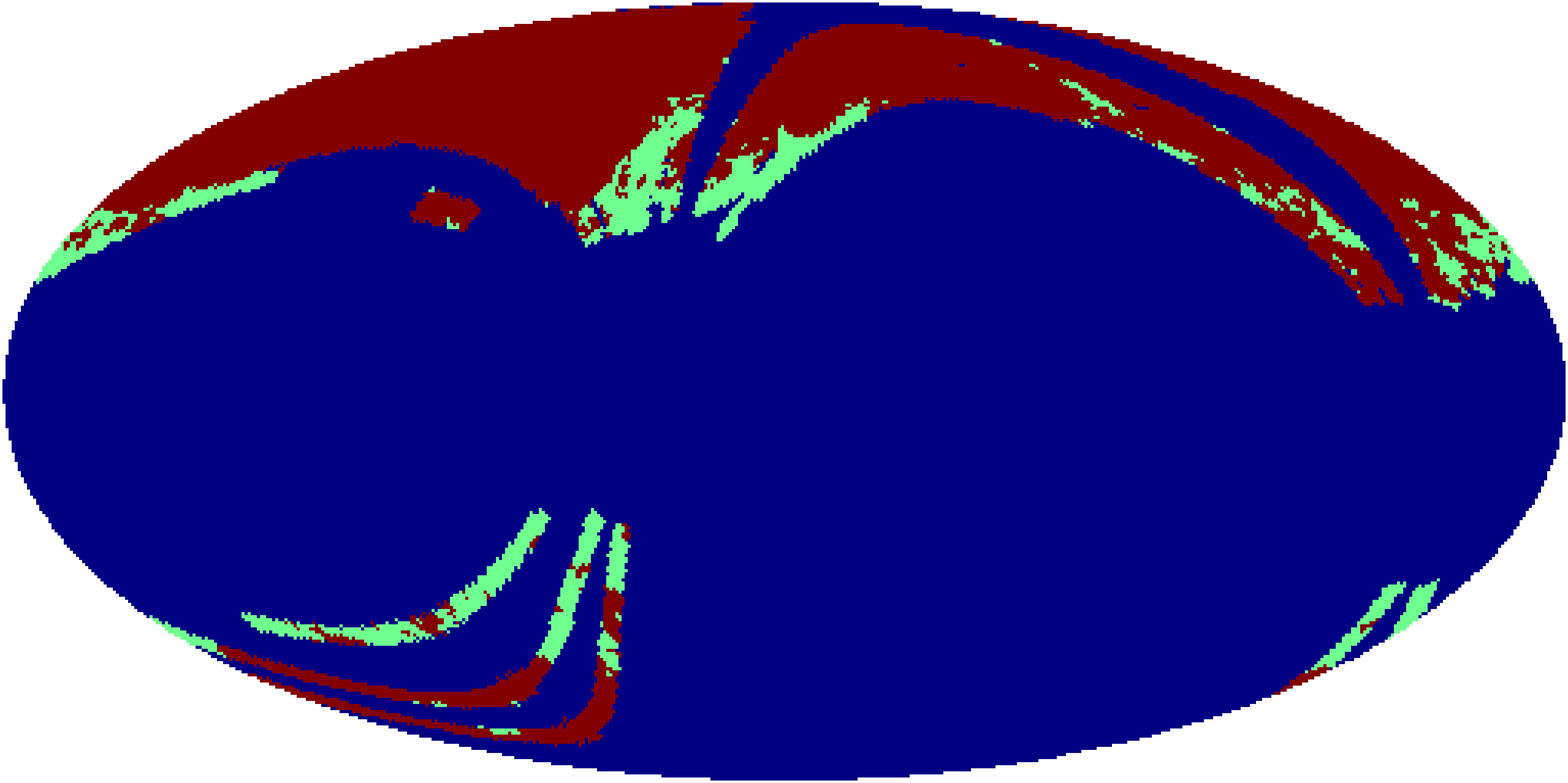}
\end{minipage}
&
\begin{minipage}{0.5\linewidth}
\includegraphics[angle=0,width=1.0\linewidth]{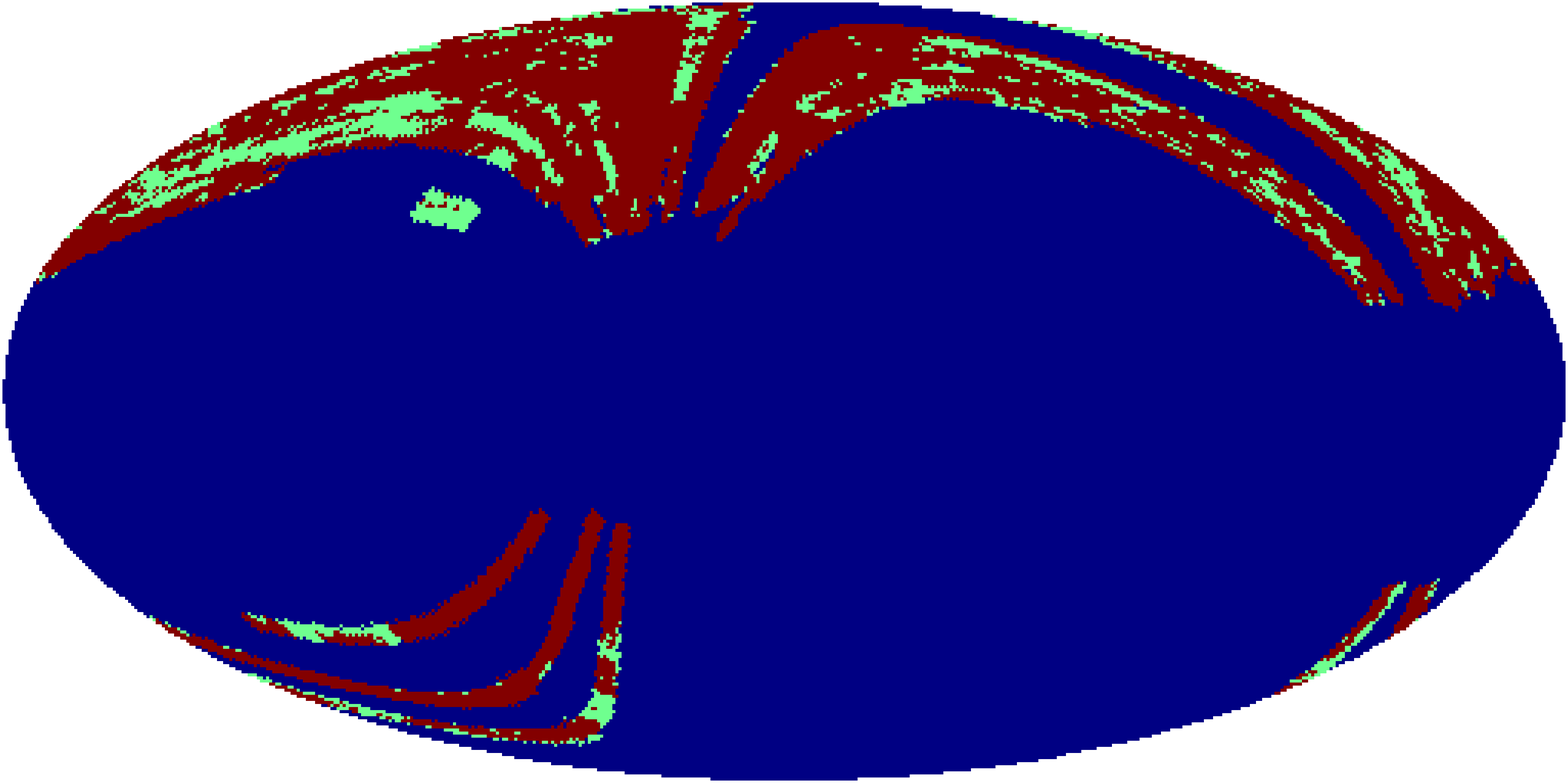}
\end{minipage}
\\
\begin{minipage}{0.5\linewidth}
\includegraphics[angle=0,width=1.0\linewidth]{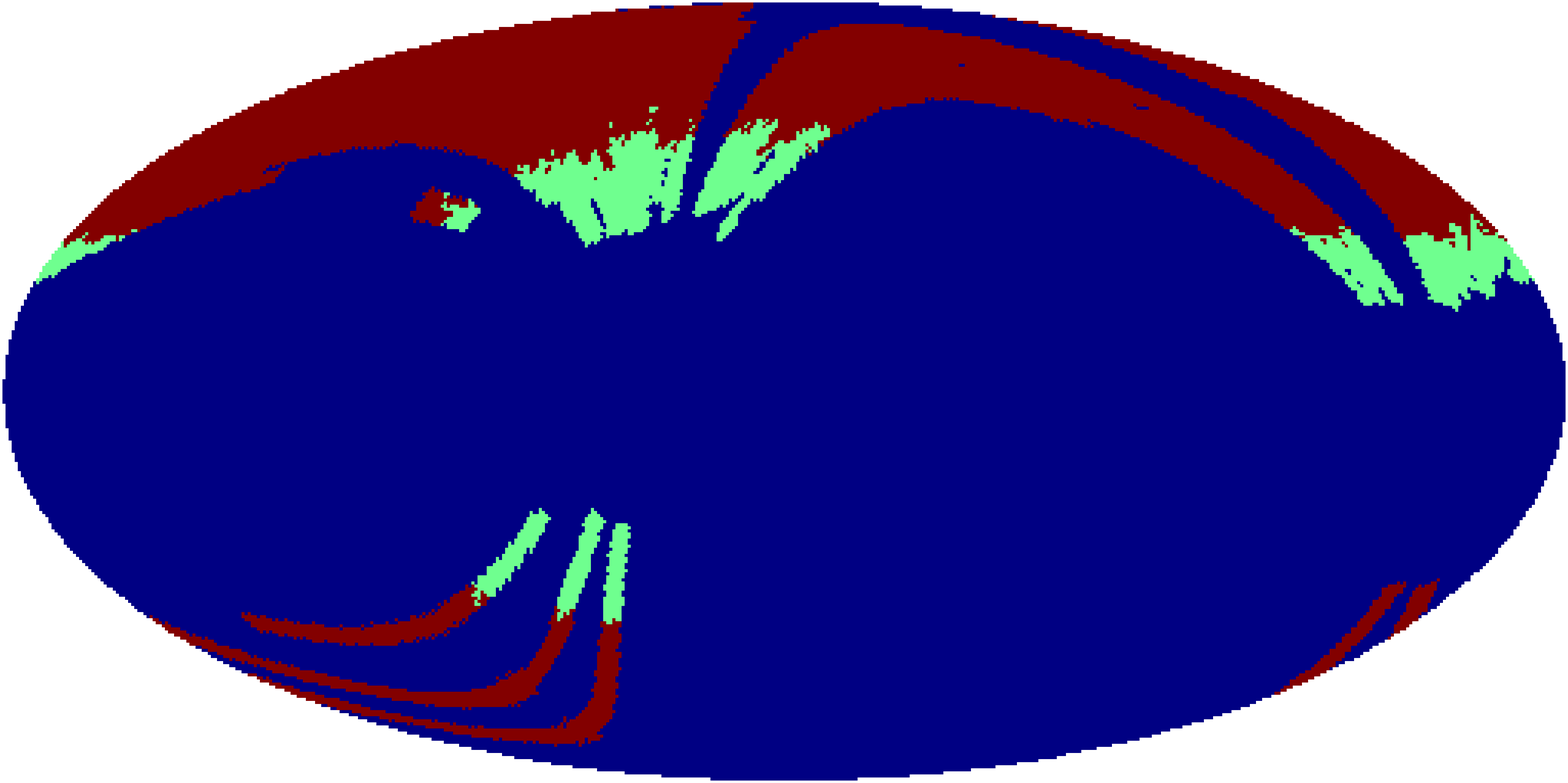}
\end{minipage}
&
\begin{minipage}{0.5\linewidth}
\includegraphics[angle=0,width=1.0\linewidth]{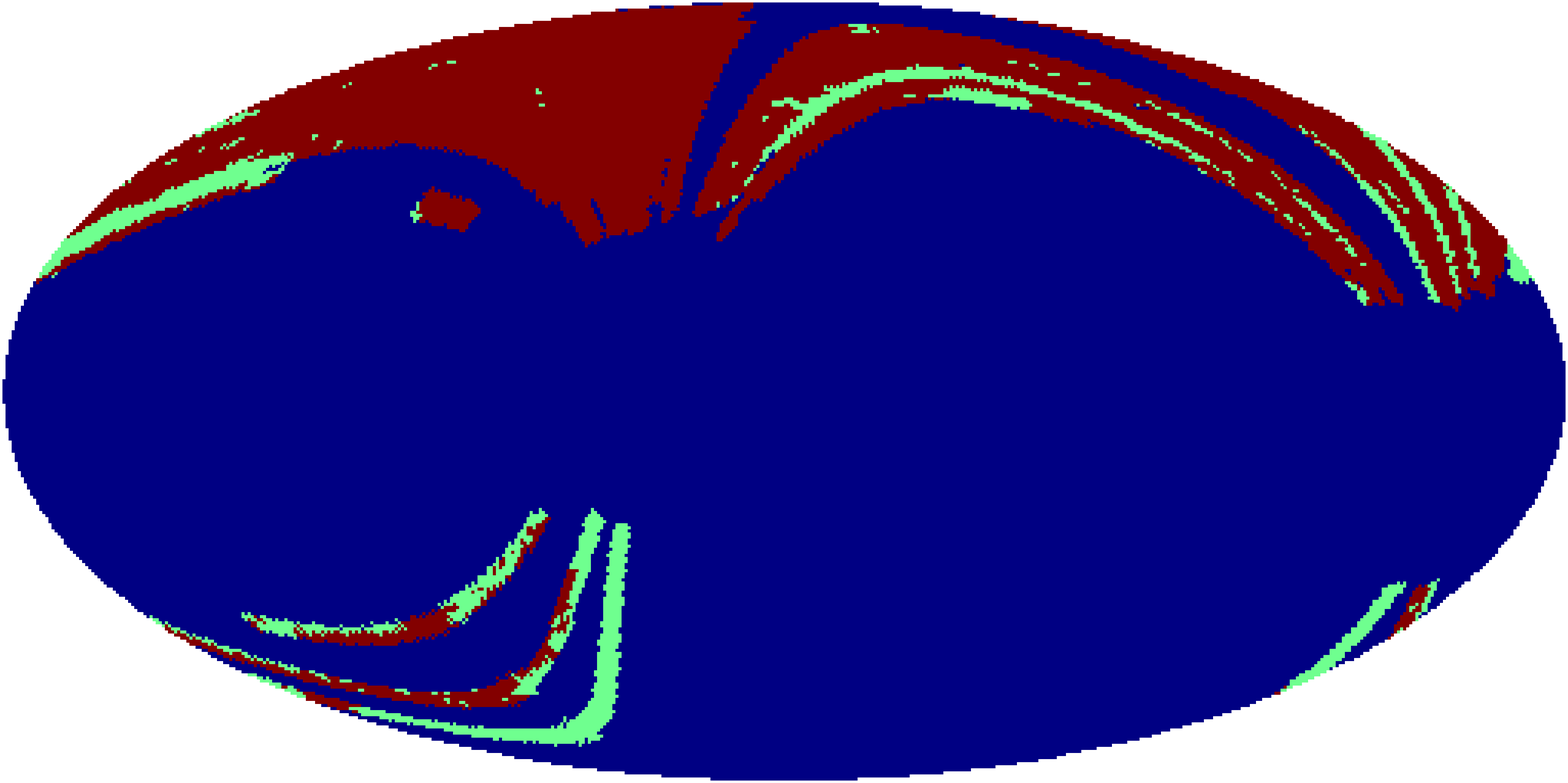}
\end{minipage}
\end{tabular}
\caption{Foregrounds masks for extinction, seeing, point sources and sky brightness in the $ r $ band. The 20\% of pixels with the worst contamination are shown in light green.  The most relevant of these effects is the extinction of the sources.}
\label {fig:fgs}
\end{center}
\end{figure}

From our random sample of SDSS galaxies, we find the value of each of these foreground quantities associated with each object (the extinction) or each  field of view (sky brightness, seeing and number of point sources), and we build their distribution in each pixel. Then we take the median and we find the distributions of the medians of all pixels. Finally, we produce the masks for each foreground excluding the worst $ 20 \% $ pixels, i.e. the pixels whose median value for a given foreground is in the upper $ 20 \% $ tail of the distribution of the medians of that foreground. These masks are shown in Fig. \ref {fig:fgs} for the $r$ band.


\section {The auto--correlation} \label {sec:acf}

To check the consistency of our method and to probe how biased our quasar sample relative to the underlying dark matter, it is useful to measure first the autocorrelation function (ACF) of the quasar catalog. To do this, we use the estimator $ \hat c^{tt} $, where the index $ tt $ refers to the total catalog (including possible contaminations):
\be
\hat c^{tt} (\vartheta) = \frac {1} {N_{\vartheta}} \sum_{i,j} f_i f_j \left(\frac{n_i}{f_i} - \bar n\right) \left(\frac{n_j}{f_j} - \bar n\right),
\ee
where the sum runs over all the pixels with a given angular separation.  As defined above, $ f_i $ is the $ i $-th pixel coverage fraction, $ n_i $ is the number of sources in the $ i $-th pixel, $ \bar n $ is the expectation value for the number of objects in the pixel.  For each angular bin centered around $\vartheta$, 
\be 
 N_{\vartheta} = \sum_{i,j} f_i f_j
\ee  
is the number of pixels pairs separated by an angle within the bin, weighted with the coverage fractions. 

\begin{figure}[htbp] 
\begin{center}
\includegraphics[angle=0,width=1.\linewidth]{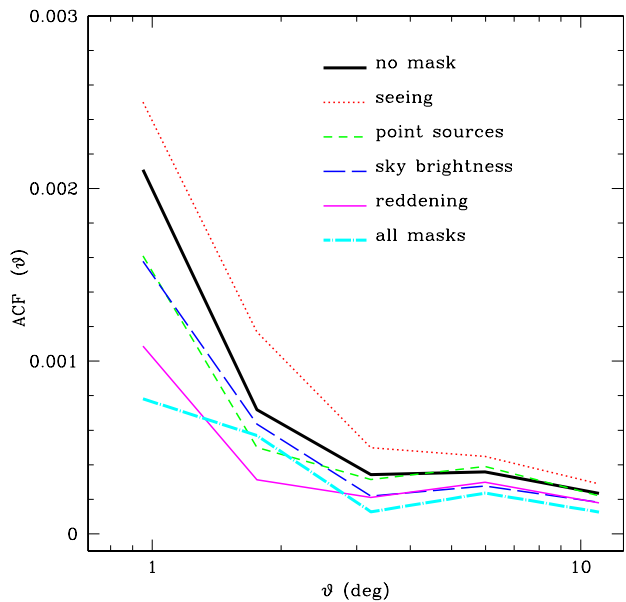}
\caption{Auto--correlation function of the quasars measured for all the sample, for a single foreground mask and for all masks joint. A similar result is obtained for $ g $ band masks.}
\label {fig:acfmasks}
\end{center}
\end{figure}

Here we present results using $ N_b = 5 $ bins of $ \vartheta $, in the range $ 0.5^{\circ} < \vartheta < 10^{\circ}$.
We tried various angular binning schemes and the results seem fairly independent assuming a sufficient number of bins are used.  Fig. \ref {fig:acfmasks} shows the ACF with and without the $r$ band based foreground masks of Fig. \ref {fig:fgs},  and the results are very similar using  $ g $ band based masks. We find that the dominant effect is the extinction; the result obtained with this mask is close to the one given by the application of all masks together.  This removal of the areas with the highest 20\% of the extinction values is equivalent to cutting pixels with a reddening in the $ g $ band $ A_g > 0.18 $, which is effectively what was done by \cite {Myers:2005jk}; for these reasons, we will use the reddening mask and not the others, in order not to excessively reduce the sample.  We have also checked that a stricter cut in reddening (30\%) does not change the result. For the other masks the 20\% threshold is likely much more aggressive than required, but the independence of the cross--correlation function (henceforth CCF) on these cuts shows that they are not significant contaminants.

\begin{figure}[ht] 
\begin{center}
\includegraphics[angle=0,width=1.0\linewidth]{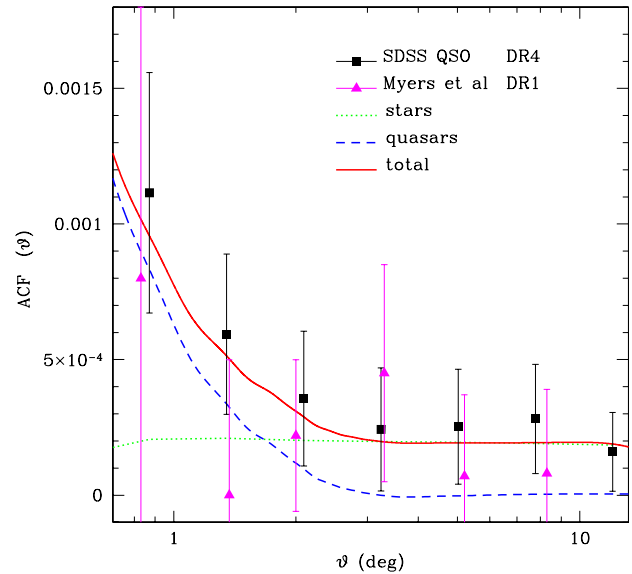}
\caption{The auto--correlation function of the quasar catalog with the reddening mask. The square (black) points are the observations $ \hat c^{tt} $, the dashed line is the expectation $ c^{tt} $, and the solid and pointed lines are its component (theoretical quasars and stellar contamination). We plot also the last points (red triangles) of the ACF measured by \cite {Myers:2005jk} for comparison.}
\label {fig:acf}
\end{center}
\end{figure}

This detection is consistent with the previous measurements \cite {Myers:2005jk, Croom:2004eg,Porciani:2004vi,Myers:2006};  these previous results used smaller data sets and were focused on smaller angular scales.  In Fig. \ref{fig:acf} we directly compare, using a similar binning, our detection to that of \cite{Myers:2005jk} which analyzed 80,000 objects from SDSS DR1 photometric catalog.

We can model the total theoretical ACF $ c^{tt} (\vartheta) $ as composed by the quasar and the star ACFs, $ c^{qq} (\vartheta) $ and $ c^{ss} (\vartheta) $, in the form
\be 
c^{tt} (\vartheta) = (1 - k)^2 c^{qq} (\vartheta) + k^2 c^{ss} (\vartheta),
\ee \label {eq:compo}
where $ k $ is the fraction of stellar contamination and we assume there is not any cross--term, due to the independence of stars and quasars (see \cite {Myers:2005jk}.) We obtain the stellar $ c^{ss} (\vartheta) $ from the average of 1000 subsamples of $ k N_q $ stars (the number of stars we expect to have in the catalog) from a random sample of $ 2 \cdot 10^6 $ stars from the SDSS survey DR4 catalog; the quasar $ c^{qq} (\vartheta) = b^2 c^{mm} (\vartheta) $ is calculated from the matter power spectrum for the best fit \WMAP third year model (\WMAP3), produced with \CMBFAST \cite {Seljak:1996is} with a given source redshift distribution and assuming a linear bias factor, $ b $ relating the quasar clustering to the matter distribution.  We have also to take in account the window function $ w (\vartheta) $ associated with our pixelization, that is given by the \HEALPIX team: the theoretical ACF $ c^{qq} (\vartheta) $ is convolved with the window function $ w (\vartheta) $.  The best values for the parameters are $ k = 0.05 \pm 0.01 $ and $ b = 2.3 \pm 0.2 $. The stellar contamination is thus in agreement with the expected value and both the stellar contamination and bias are consistent with 
those measured by \cite {Myers:2005jk,Myers:2006}.

We have calculated the errors on the total ACF shown in Fig. \ref {fig:acf} by producing 1000 random quasar maps, with the same statistics as the total catalog and an added Poisson noise.


\section {The cross--correlation} \label {sec:corr}

For the cross correlation analysis, we use the \WMAP \emph {Internal Linear Combination} (ILC) map derived from the third year \WMAP data \cite {Hinshaw:2006ia}, pixelized in the same way and with the same resolution as the quasar map.  Even though this ILC map was already built to minimize the Galactic and other foreground contaminations, we have applied to it the most severe mask given by \WMAP, the \emph {kp0} mask, which corresponds to a cut of $ 32 \% $ of the sky.  We checked that the results do not change significantly if we use the different frequency band maps V and W, corresponding respectively to 61 and 94 GHz, all with the same masking; the results change slightly using the Q band map, which is the most affected by Galactic synchrotron contamination (see section \ref {sec:corr}). 
We have also checked that the result remains consistent using the \WMAP 1st year ILC map, and also does not depend on  whether we use the smoothed or the raw single band maps.

\begin{figure}[htbp] 
\begin{center}
\includegraphics[angle=0,width=1.\linewidth]{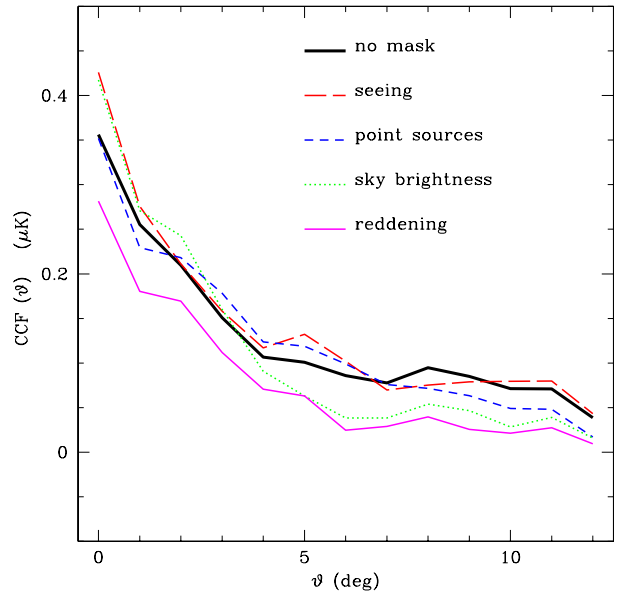}
\caption{Cross--correlation function of the quasars and the CMB measured for various foreground masks. Similar results are obtained for $ g $ band masks.}
\label {fig:ccfmasks}
\end{center}
\end{figure}

To measure the cross--correlation function (CCF) between the quasar map and the \WMAP ILC map, we used the estimator
\be
\hat c^{Tt} (\vartheta) = \frac {1} {N_{\vartheta}} \sum_{i,j} f_i \left(T_j - \bar T\right) \left(\frac{n_i}{f_i} - \bar n\right),
\ee
where  $ T_j $ is the CMB temperature in the $ j $-th pixel and $ \bar T $ is the expectation value for the CMB temperature respectively. We again down--weight the partially filled pixels and $N_{\vartheta} $ is defined as above, but with a single weighting factor. 
We calculated this function in $ N_b = 13 $ bins of $ \vartheta $, in the range $ 0^{\circ} < \vartheta < 12^{\circ} $, with and without using the foreground masks of Fig. \ref {fig:fgs}, obtaining the results shown in Fig. \ref {fig:ccfmasks}.  We obtain very similar results using the $ r $ and $ g $ band masks.  The reddening mask is the one that yields the lowest CCF; to be conservative and consistent with the ACF measure, we choose to apply this same mask.  As expected, however, the reddening dependence is weaker for the cross--correlation measurement than for the quasar ACF. 

\begin{figure}[ht] 
\begin{center}
\includegraphics[angle=-0,width=1.0\linewidth]{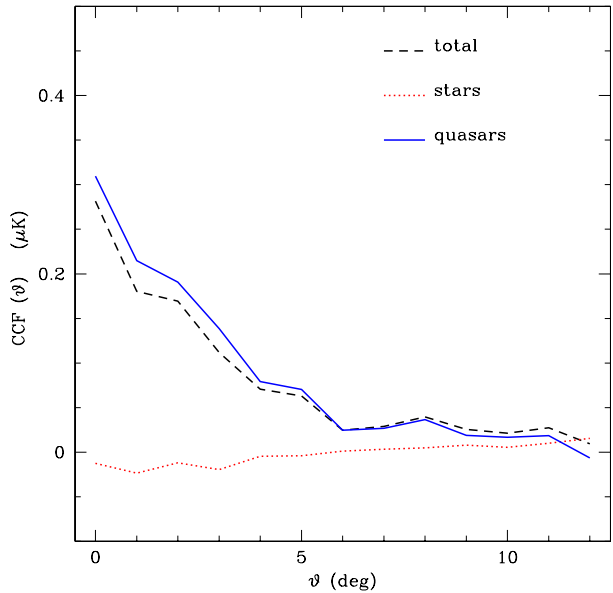}
\caption{The cross--correlation with the quasar catalog shows a small stellar contamination in the \WMAP 3 ILC map with the \emph{kp0} mask. The dashed line is the measured total CCF $ \hat c^{Tt} $, while the point--dashed line is the measured stellar CCF, $ \hat c^{Ts} $. The solid line is the difference between the two, which is our estimator for the true quasar CCF, $ \hat c^{Tq} $.}
\label {fig:starcont}
\end{center}
\end{figure}

Fig. \ref{fig:starcont} displays the CCF between the WMAP3 ILC map and our NBC--KDE quasar sample. In reality, this is a measure of the cross--correlation between the CMB and a mixed sample of quasars and stars: although one does not expect a correlation between the cosmic radiation and local stars, we measured a small but non zero result. This indicates that the \WMAP3 ILC map, even after the most severe \emph {kp0} masking, still has a small residual Galactic contamination. The stellar correlation has to be subtracted from the total detection yielding
\be \label{eq:estqs}
\hat c^{Tq} (\vartheta) = \frac {\hat c^{Tt} (\vartheta) - k \hat c^{Ts} (\vartheta)} {1 - k}.
\ee
We compare this to the theoretical expected function $ c^{Tq} (\vartheta) $ calculated again from the \LCDM model with the \WMAP3 best fit parameters, using a program based on \CMBFAST \cite{Seljak:1996is}.

To estimate the errors on the CCF and the covariance matrix, we use three different Monte Carlo methods.   The first method is to produce a high number (2000) of random CMB maps with the \WMAP best fit parameters and cross--correlate them with the true quasar map, after the application of the same \emph {kp0} mask.  Alternatively, we can do the reverse: use the true temperature map and make random maps of the quasars using the \WMAP parameters and our observed bias, to which we add  
the Poisson error on the counts in the pixels.  Both of these approaches produce similar answers, showing the covariances seem to be independent of any peculiarities of either of the two observed maps.  

These approaches give the covariances assuming the absence of correlations; while this should work well assuming any true correlations are weak, it is important to understand the extent to which the presence of correlations will bias the covariance calculation.  Indeed, if there are strong correlations, then these approaches should overestimate the errors.
To account for correlations, we want to generate random temperature and quasar maps with the same ACF and CCF of the measured maps, including also the Poisson uncertainty in the quasar counts.  

Based on the standard \LCDM model, we can generate the expected angular power spectra for the anisotropies $ C_l^{TT}, C_l^{Tq}, C_l^{qq} $ for the temperature only, the cross--correlation and the quasar autocorrelation.  Here, the cross spectrum is assumed to arise solely from the ISW effect.    From these power spectra, we can generate three random maps and use them to calculate the errors in the cross--correlation \cite{Boughn:1997vs}.
We begin by making random temperature maps, $ T_i^r $, based on $ C_l^{TT} $.  (We neglect any noise which is thought to be small on the scales of interest.) We then decompose the quasar power spectrum into two parts:
\be
C_l^{qq} \equiv C_l^{qq \parallel} + C_l^{qq \perp},
\ee
where the parallel and orthogonal signs indicate completely correlated and uncorrelated with respect to the temperature map, and
\bea
C_l^{qq \parallel} & \equiv & \frac {(C_l^{Tq})^2} {C_l^{TT}} \nonumber \\
C_l^{qq \perp}     & \equiv & C_l^{qq} - \frac {(C_l^{Tq})^2} {C_l^{TT}}.
\eea
Using $ C_l^{qq \parallel} $ and the same phases as for the temperature map, we create a correlated quasar density map, $ \delta_i^{r \parallel} $; we add to this an uncorrelated quasar density map,$ \delta_i^{r \perp} $ created using $ C_l^{qq \perp} $, with independent random phases. 
The total quasar density is $ \delta_i^r = \delta_i^{r \parallel} + \delta_i^{r \perp} $, and we can now build a random total quasar map $ n^r_i $, as
\be
n_i^r = (1 + \delta_i) \bar n.
\ee
Finally, we can add random Poisson noise to this, which we derive from the quasar number in each pixel. 

Generating 2000 Monte Carlo simulations $ n_i^r $ and correlating them with the random temperature map  $ T_i^r $ we can now find the covariance matrix due to sample variance, $ R_{ij}^{samp} $.  The results are consistent with what we obtain with the errors in the temperature only.

\begin{figure}[htb] 
\begin{center}
\includegraphics[angle=0,width=1.0\linewidth]{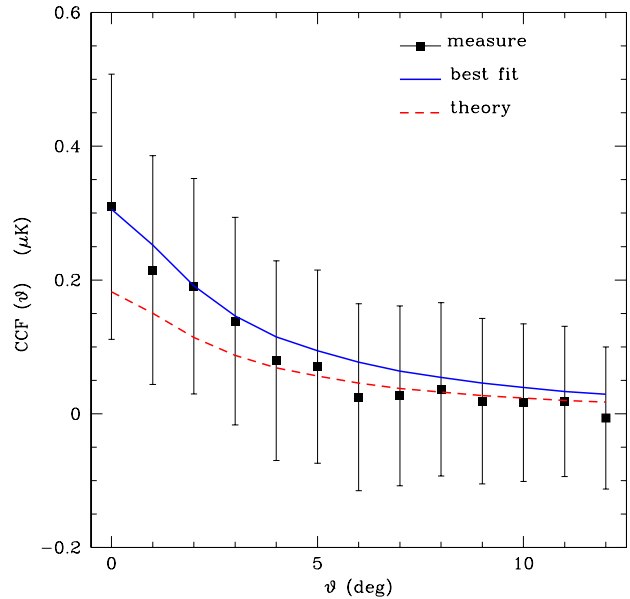}
\caption{The measure of the cross--correlation. The points are the observed correlation between WMAP and the quasars, the solid line is the best fit  $ \Lambda $CDM theoretical model and the dashed line is the prediction for the \WMAP3 best fit model with $ b = 2.3 $. The points are highly correlated: the typical level of correlation between two neighbouring bins is $ \sim 95 \% $.}
\label {fig:ccf}
\end{center}
\end{figure}

The errors in the estimate made from Eq. (\ref {eq:estqs}) should also include measurement errors; assuming the mask is known, these can arise from uncertainties in $k$, the fraction of stellar contamination, or from the assumed stellar cross correlation $c^{Ts}$.  The full covariance is thus approximately:  
\be \label {eq:RTq}
R_{ij}^{Tq} \simeq  R_{ij}^{samp} + k^2 R_{ij}^{Ts} + \gamma_{ij} \sigma_k^2,
\ee
where 
\bea
\gamma_{ij} & = &  (\hat c^{Tt}_i - \hat c^{Ts}_i) (\hat c^{Tt}_j - \hat c^{Ts}_j) 
\eea
and we have assumed the stellar contamination $k \ll 1$ and $ \sigma_k $ is its error.
We account for the uncertainty of the star CCF $ R_{ij}^{Ts} $ calculating cross--correlations between random samples of $ k N_q $ stars and the CMB map.  The best fit and diagonal errors are shown in Fig. \ref {fig:ccf}.

Three of the points in Fig. \ref {fig:ccf} are more than $ 1 \sigma $ greater than zero, but the points are all highly correlated: the total significance of the ISW detection is found to be conservatively $ 2.1 \sigma $.  Making less conservative choices in the analysis can lead to slightly higher significance, but it does not exceed  $ 2.5 \sigma $. 
This significance is based on a theoretical model for the expected ISW signal; using a modified version of the \CMBFAST code, we calculated the predicted cross--correlation function for the \WMAP3 best fit $ \Lambda $CDM model, for a matter map with the redshift selection function shown in Fig. \ref {fig:zeta}. We include the pixel window function and bin the expected correlation into the $ N_b = 13 $ $ c_i $ in the same way as we calculated the experimental CCF.

We can compare the theoretical CCF $ c_i $ (the index $ Tq $ is understood) with the observed values $ \hat c_i $ and assume a Gaussian likelihood model as
\be
\mathcal {L} = (2 \pi)^{-N / 2} [\det R_{ij}]^{-1 / 2} \exp [- \sum_{ij} R_{ij}^{-1} (\hat c_i - c_i) (\hat c_j - c_j)/2],
\ee
where $ R_{ij} $ is the CMB--quasar cross--correlation function covariance matrix defined in Eq. (\ref{eq:RTq}).  For a given distribution of sources, the shape of the theoretical curves remains unchanged to a good approximation; however, the amplitude of the cross correlation strongly depends on the cosmological model, so that we can write \cite{Boughn:1997vs}
\be
c^{Tq} (\vartheta) = A(\Omega_m,w) g^{Tq} (\vartheta),
\ee
where $ g^{Tq} (\vartheta) $ is normalized to 1 at $ 0^{\circ} $.
We found the best value for $ A  $ maximizing the likelihood, i.e.
\be
A = \frac {\sum_{i,j=1}^N {R_{ij}^{-1} g_i \hat c_j } }{\sum_{i,j=1}^N R_{ij}^{-1} g_i g_j},
\ee
and the variance
\be
\sigma^2_A = {\left[ \sum_{i,j=1}^N R_{ij}^{-1} g_i g_j  \right]^{-1}}.
\ee
The best fit for the CCF is $ A = (0.30 \pm 0.14) \mu $K.

\begin{figure}[htb] 
\begin{center}
  \includegraphics[angle=0,width=1.0\linewidth]{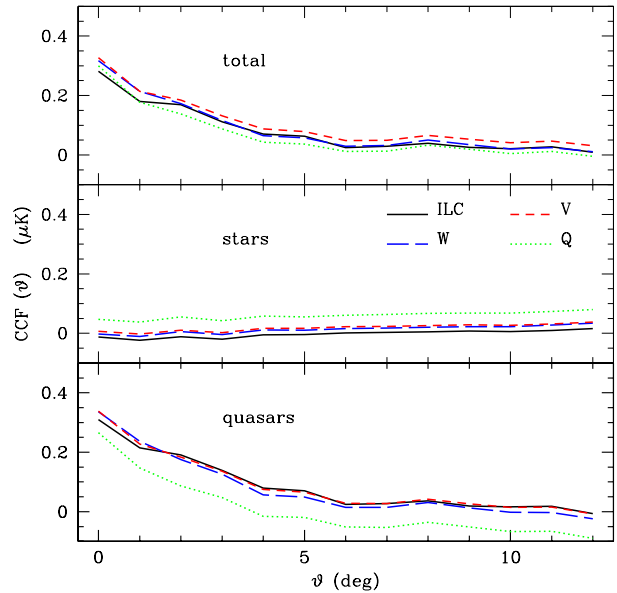}
\caption{Frequency dependence of the cross--correlation, for the ILC map (black, solid), W (blue, long dashed), V (red, short dashed) and Q (green, dotted) bands. The first panel shows the measured CCF with the reddening corrected KDE catalog; in the second panel we plot the observed CCF with our random star samples (see text); the last panel shows the subtraction of the stellar contamination.}
\label {fig:multi}
\end{center}
\end{figure}

In order to study the frequency dependence of our result, we measure the cross--correlation using the single band CMB maps (W, V and Q); in the first panel of Fig. \ref {fig:multi} we show the measured CCF for the single band maps, and we can see that the measure is almost frequency independent. We see in the second panel that the correlation with the random star sample is different for the different maps, being low as expected for the ILC map, increasing for the W and V bands, in which the Galactic contamination is more relevant, and being even bigger for the Q band, which is significantly affected by synchrotron radiation. The effect of the subtraction is shown in the last panel: the resulting CCF is still frequency independent and very consistent for the ILC and the V and W bands, while the galactic contamination starts to be important in the Q band, actually hindering the measure of the cross--correlation for this band. 


\section {Cosmological constraints} \label {sec:resul}

To compare with these observations, we calculate the expected ISW cross--correlations based on linear theory using a modified version of the \CMBFAST code \cite {Corasaniti:2005pq}. We calculate the quasar auto--correlations in the same way and assume a non--evolving linear bias factor.  For the purposes of calculating the expected cross correlation, we use the actual measured redshift selection function of the sample $ \varphi = dN / dz $ normalized to unity.  Were the quasar bias to evolve with redshift (e.g., as detected by \cite{Porciani:2004vi,Croom:2004eg,Myers:2005jk,Myers:2006}), this would effectively shift the redshift weighting.   We use the measurements of the quasar auto--correlation function to determine this bias to be $b = 2.3 \pm 0.2$, consistent with previous measurements made at smaller scales \cite{Porciani:2004vi,Croom:2004eg,Myers:2005jk,Myers:2006}.     
 
Assuming the cross--correlation that we see is due to the ISW effect, we can put some constraints on the nature of dark energy.   First, consider the pure CDM model without any dark energy, ($ \Omega_m = 1, \Omega_{\Lambda} = 0 $);  by relaxing some assumptions, such as using a strongly broken power law for the primordial power spectrum, such models might be consistent with the \WMAP data \cite{Blanchard:2003du}.  However, these models would predict no ISW correlations, so would be disfavored at the $ 2 \sigma $ level with this data alone, and even more strongly when other ISW observations are included.    

\begin{figure}[ht] 
\includegraphics[angle=0,width=\linewidth]{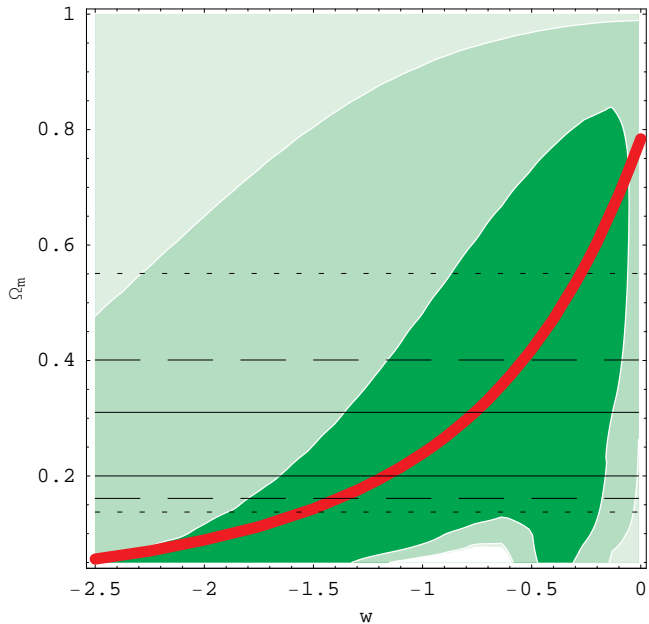}
\caption{Likelihood contours within 1, 2 and 3 $ \sigma $ on $ w - \Omega_m $ inferred from this ISW detection alone. On the thick red line lies the family of the models whose TT spectrum matches the \WMAP measured one, having the right comoving distance to the LSS, according to Eq. (\ref {eq:peak}). The thin black lines are the constraints on the Hubble parameter $ h $ \cite{Freedman:2000cf} at 1, 2 and 3 $ \sigma $ (solid, dashed and dotted) assuming $\omega_m = 0.128$.}
\label {fig:WO}
\end{figure}

Next, consider a flat dark energy dominated model (wCDM) with constant equation of state $w$.   We first explore the likelihood function of the parameters $ \Omega_m, w $ with the constraint that the values of $ \omega_b \equiv \Omega_b h^2, \omega_m \equiv \Omega_m h^2 $ and the other parameters are fixed to the \WMAP3 best fit values ($ \omega_b = 0.0223, \omega_m = 0.128$) \cite{Spergel:2006hy}.  Here and below, the Hubble parameter is  $100 h {\rm km}s^{-1}{\rm Mpc}^{-1}$. We obtain the result shown in Fig. \ref {fig:WO}.

We can see that the data are in favor of the \LCDM model, but due to the weak detection only models far away from this are actually ruled out. Most notably, models with a very small matter fraction predict too large a correlation and are inconsistent with the measurement.  Fixing $ w = -1 $ yields the $ 1\sigma $ interval for the matter fraction $ 0.075 \le \Omega_m \le 0.475 $.

While we have used \WMAP constraints on the matter and baryon densities above, most models will actually be inconsistent with the positions of the CMB Doppler peaks.   It is interesting to consider the family of models which are consistent with the full temperature power spectrum measurements: for this, we need the angular scale of the Doppler features to be fixed to the observations.   The sound horizon scale is effectively fixed when we fix $\omega_m$ and $\omega_b$, so we must add the additional constraint that the models have the same comoving distance to the last scattering surface, $ D^A_* $, given by
\be \label {eq:peak}
D^A_* =  \frac {1} {H_0} \int_{1/(1+z_*)}^1 \frac {1} {\sqrt{\Omega_{\Lambda} a^{4-3(1+w)} + \Omega_m a}} da,
\ee
where $ z_* $ is the redshift of the last scattering surface, weakly dependent on $ \omega_m $.     
$ D^A_* $ is kept constant if the variations in $ w $ are compensated by changes in the Hubble parameter $ h $ and the matter density $ \Omega_m$: in Fig. \ref {fig:WO} we show the family of the models fulfilling this condition, and we see that most of them are compatible with our ISW detection.

\begin{figure}[ht] 
\begin{center}
\includegraphics[angle=0,width=\linewidth]{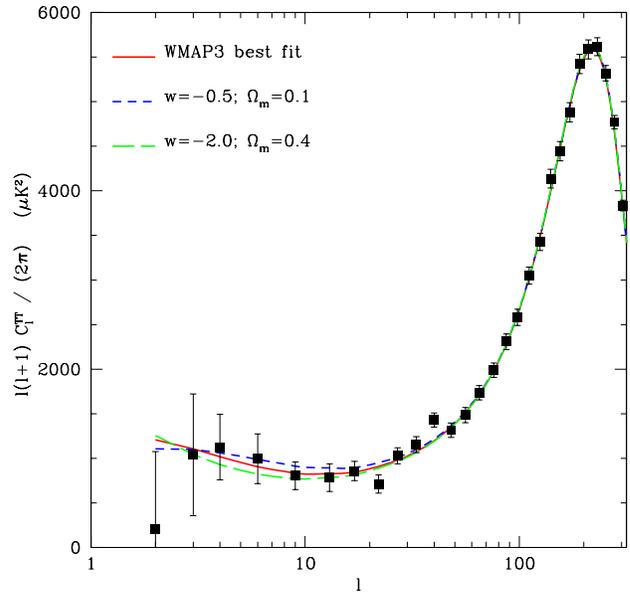}
\caption {Temperature power spectrum of models with the same angular diameter distance to the last scattering surface as the \WMAP3 best fit compared with the binned \WMAP3 data (points). Changes in $ w $ are compensated by changes in $ \Omega_m $, but $ h $ undergoes considerable variations.}
\label {fig:TT}
\end{center}
\end{figure}

This range of models is consistent both with the CMB autocorrelation and cross--correlation measurements: for instance, we show in Fig. \ref {fig:TT} the temperature power spectra of two of these models; it is slightly different from the \WMAP 3 best fit only at very large scales.

Note however that many of these models are inconsistent with direct measurements of the Hubble constant.  For $w = -0.5$, the Hubble constant would have to be as low as $h = 0.55$, while for $w=-2.0 $ the Hubble constant would be unreasonably high, $h = 1.20$.   Current limits on the Hubble constant, e.g. $ h = 72 \pm 8 $ \cite{Freedman:2000cf}, would constrain our measured $ w $ in the range $ -1.18 \le w \le -0.76 $.  Models with $w$ in the range are practically indistinguishable from the best fit cosmological constant model plotted in Fig. \ref {fig:gazplot}.

We have also investigated different classes of models which might be more likely to produce a significantly different ISW signal at the quasar redshift.  There are two ways models might be strongly ruled out given our relatively weak detection: either they predict a correlation of the opposite sign, or they predict a much higher amplitude of correlation.  Producing a negative correlation requires that the gravitational potential grow in time rather than decay, which is difficult to arrange in typical dark energy scenarios because the accelerated expansion tends to slow down the growth of structure; one possibility is a closed model without dark energy, as suggested by Nolta {\it et al.} \cite{Nolta:2003uy}.   

Producing a much larger signal at high redshifts is also difficult, given the other constraints on dark energy.   For the models discussed above where dark energy scales as a power law, its fractional density tends to be small at a redshift of $z \sim 1.5$.  In a cosmological constant model, $\Omega_{DE}(z=1.5) = 0.16$; while this density can be higher for $w > -1$, the transition to dark energy domination becomes less sudden, leading to a smaller effect.  If however there were a sharp drop in the dark energy density, it would be possible to be for the dark energy to be large at high redshifts while still remaining compatible with constraints from lower redshifts.   In such a model, our measurement can limit the dark energy density at $z = 1.5$;  models with  $\Omega_{DE}(z=1.5) < 0.5$ would produce a much higher cross correlation than is observed, and can be ruled out at the $3\sigma$ level.    

An alternative explanation of the dark energy problem is to modify the laws of gravity on large scales; such theories may have consequences for structure formation which are significantly different to dark energy models, even if the background expansion appears the same.   The ISW correlations are an important way of probing these differences, particularly at high redshift \cite{Lue:2003ky}.  There are many ways of implementing such changes, but much recent work has focused on the extra--dimensional model of Dvali, Gabadadze and Porrati (DGP) \cite{Dvali:2000hr, Koyama:2005kd}; under some assumptions, Song, Sawick and Hu \cite{Song:2006jk} have recently shown the ISW signal to be comparable to dark energy at low redshift, but significantly higher above $z=1$ in an open DGP model.  At a redshift of $z=1.5$, this enhancement increases the expected cross correlation by a factor of two; while still consistent with our present observations, larger quasar samples could be used to constrain such models in the future.   

\subsection* {Comparison of ISW detections}

\begin{figure}[ht] 
\begin{center}
  \includegraphics[angle=0,width=1.0\linewidth]{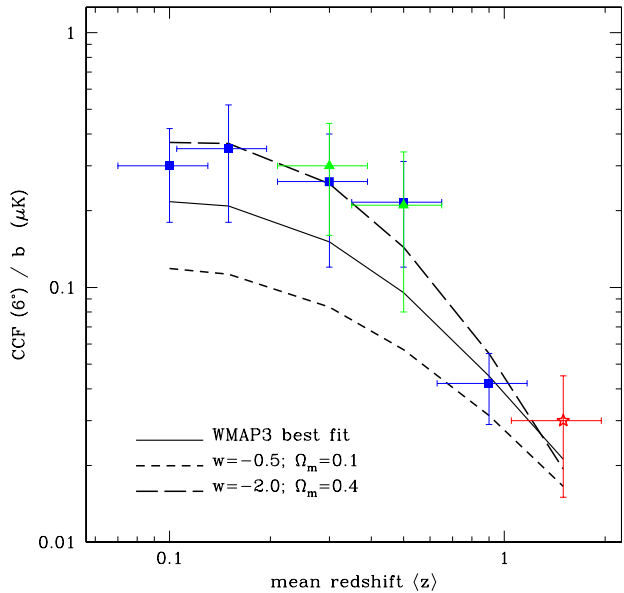}
\caption{Summary of the detections of the ISW effect through cross--correlation with different catalogs, compared with \WMAP3 best fit model. The blue (squared) points are in the order the correlations with 2MASS, APM, SDSS, SDSS high-z, NVSS+HEAO, as collected by \cite {Gaztanaga:2004sk}; the green (triangular) points are the measure by \cite{Cabre:2006qm}, while the red (star) point is our KDE--QSO measure. The lines are the theoretical expectations for \WMAP3 best fit model (solid), and two models with $ w = -2 $ (long dashed) and  $ w = -0.5 $ (short dashed) respectively (see text for details of the models).}
\label {fig:gazplot}
\end{center}
\end{figure}

We can now compare our detection with previous ones.  Following previous convention \cite{Gaztanaga:2004sk, Corasaniti:2005pq}, we plot the observed CCF at $ 6^{\circ} $ in function of the mean redshift of the survey in Fig. \ref {fig:gazplot}.  This angular scale is chosen to avoid possible contamination from other effects that are dominant on smaller scales, such as lensing and SZ; however, it should be remembered that this representation is a one dimensional slice of the correlation function data.  This approach can suppress the high redshift measurements, where a given angle corresponds to a larger physical scale.  In the figure, we also add a conservative $ 30 \% $ error on the estimation of the mean redshift of the surveys.

We also plot the theoretical expected values for the CCF at these redshifts for the models of Fig. \ref {fig:TT}. As above, we assume consistency with the CMB power spectrum, i.e. we fix the comoving distance to the last scattering surface defined in Eq. (\ref {eq:peak}), while other parameters are fixed to the best fit \WMAP3 values.
We see that the behavior is largely that expected from a cosmological constant model, with the amplitude dropping off at high redshifts.  While many of the measurements are actually higher than expected, the differences are largely within the expected errors.  This provides further support that the observed cross--correlations are due to the ISW effect.   

\section {Conclusions} \label {sec:concl}

Here we have presented evidence of a weak correlation between the CMB and the distribution of high redshift 
quasars detected in the SDSS.  Its amplitude, angular dependence and independence of the CMB frequency are all consistent with the interpretation as due to the integrated Sachs--Wolfe effect, with a significance in the range $2-2.5 \sigma$, robust to changes in the mask and assumptions about stellar contamination.  

Without dark energy, no such correlation is expected.  With a mean quasar redshift of $z = 1.5$, this represents the earliest evidence yet for dark energy and gives us a means to further probe its evolution.  Our measurements directly limit the density of dark energy at high redshifts, independent of its lower redshift behavior.  They can also potentially provide interesting limits on alternative models with modified gravity.   

These measurements will be improved when the photometrically classified quasar data set is extended to the entire SDSS area.  With a data set 40\% larger, the photometric redshifts could be used to split the sample into two broad redshift bins, above and below $z = 1.5$, potentially allowing the evolution of the ISW effect to be seen within one self-consistent sample.  To obtain even stronger cosmological constraints, all the various ISW measurements should be combined carefully, including possible covariances which arise from the overlap of the different surveys in sky coverage and redshift; we are presently pursuing this \cite{Scranton:2006} with an aim to provide an independent probe of the nature and evolution of dark energy.  

\subsection*{Acknowledgments} 
We thank Bj{\"o}rn Malte Sch{\"a}fer and Jussi V{\"a}liviita for useful conversations.  We acknowledge support by a grant from PPARC.

Funding for the SDSS and SDSS-II has been provided by the Alfred P. Sloan Foundation, the Participating Institutions, the National Science Foundation, the U.S. Department of Energy, the National Aeronautics and Space Administration, the Japanese Monbukagakusho, the Max Planck Society, and the Higher Education Funding Council for England. The SDSS Web Site is \texttt {http://www.sdss.org/}.

The SDSS is managed by the Astrophysical Research Consortium for the Participating Institutions. The Participating Institutions are the American Museum of Natural History, Astrophysical Institute Potsdam, University of Basel, Cambridge University, Case Western Reserve University, University of Chicago, Drexel University, Fermilab, the Institute for Advanced Study, the Japan Participation Group, Johns Hopkins University, the Joint Institute for Nuclear Astrophysics, the Kavli Institute for Particle Astrophysics and Cosmology, the Korean Scientist Group, the Chinese Academy of Sciences (LAMOST), Los Alamos National Laboratory, the Max-Planck-Institute for Astronomy (MPIA), the Max-Planck-Institute for Astrophysics (MPA), New Mexico State University, Ohio State University, University of Pittsburgh, University of Portsmouth, Princeton University, the United States Naval Observatory, and the University of Washington. 


\end {document}